\documentclass{aastex}
\usepackage{spr-astr-addons}
\usepackage{url}

\RequirePackage{color}

\def\be{\begin{equation}}
  \def\ee{\end{equation}}
\def\bea{\begin{eqnarray}}
\def\eea{\end{eqnarray}}

\begin{document}
\title{Holographic f(T) gravity model}
\slugcomment{Not to appear in Nonlearned J., 45.}
\shorttitle{Short article title}


\author{A. Aghamohammadi$^{a}$\altaffilmark{1}\altaffilmark {$\ast$}}  
\affil{$^{a}$Sanandaj Branch, Islamic Azad University, Sanandaj, Iran.} \and


\altaffiltext{1}{a.aqamohamadi@gmail.com }
\altaffiltext{2}{a.aghamohamadi@iausdj.ac.ir }

\begin{abstract}
We try to study the corresponding relation between  $f(T)$ gravity and holographic dark energy (HDE).  A kind of  energy density  from $f(T)$ is introduced which has  the same role as the HDE density.  A $f(T)$ model according to the the HDE model is calculated .We find out a torsion scalar T based on the scalar factor is assumed by {\citep{sno}}.
    The effective torsion equation of state,  deceleration  parameter of the holographic $f(T)$- gravity model are calculated.
\end{abstract}



\section{Introduction}
In this work our aim  reconstruct a $f(T)$ modified teleparallel gravity model corresponding to the holographic dark energy (HDE) density.\\
An approach to investigate the essence of dark energy is well-known to the  HDE density that is proposed by {\citep{hsu, ht, zh, sj}} and have attracted a lot of interested recently. this density is defined a s follow
 \begin{equation} \label{1e}
\rho_{\Lambda}= 3c^2 M_p^2L^{-2},
\end{equation}
where $L$, is the size of the current universe, $c^2$ is a numerical constant of order unity, $M_p^{-2}=8\pi G$ is the reduced Planck mass and $\rho_{\Lambda}$ is considered as zero-point energy density {\citep{ml, ht, as, sj}}. As a rule,  many  choices are adopted  for the infrared cuoff of the universe, e.g, particle horizon, future event horizon and Hubble horizon {\citep{hq, ml}}, which we choose  the latter.\\
On the other hand, in modern cosmology the telleparallel Lagrangian density represented by the torsion scalar $T$ has been replaced to a function of $T$ as $f(T)${\citep{gr, ev}}, which is equivalent to the concept of $f(R)$ gravity, these can explain both inflation {\citep{rf}} and the late time accelerated expansion of the universe.

This paper is organized as  follows. In section $2$, we will review $f(T)$ gravity cosmology and general  properties of  the model. In section $3$, we  make the connection between the HDE  and $f(T)$) gravity and reconstruct a $f(T)$ model according to HDE model. We find out a torsion scalar $T$, also we calculate the effective torsion equation of state  and deceleration parameter. Section $4$. is devoted to the conclusion.
\section{Basic equations  of the $f(T)$) gravity model}
Our starting action for  The modified teleparallel  describing $f(T)$) gravity {\citep{ev,gr,jb,kk,ka,ml,pw,ph,rj,rf}} is as follows
 \begin{eqnarray}\label{1bt}
 I=\int\mathrm{d}^4x|e|\Big[\frac{f(T)}{2k^2}+L_m\Big],
  \end{eqnarray}
where $k^2=8\pi G,\,|e|=det(e^i_{\mu})=\sqrt{-g}$ and $e^i_{\mu}$ forms the tangent vector of the manifold, which is used as a dynamical object in teleparallel gravity, $L_M$ is the Lagrangian of matter. Taking the variation of action Eq. (\ref{1bt}) with respect to the vierbein $e^i_{\mu}$, in the flat FLRW background, the gravitational field equations can be written in the equivalent forms of those in general relativity as
\begin{eqnarray}\label{12}
\frac{3}{k^2}H^2&=&\rho_m+\rho_T,\cr
\frac{1}{k^2}(2\dot{H}+3H^2)&=&-(p_m+p_T)
\end{eqnarray}
where
\begin{eqnarray}\label{13}
\rho_T&=&\frac{1}{2k^2}(2Tf_T-f-T),\cr
T&=&-6H^2
\end{eqnarray}
and
\begin{equation}\label{14}
p_T=-\frac{1}{2k^2}\Big(-8\dot{H}Tf_{TT}+f_T(2T-4\dot{H})-f+4\dot{H}-T\Big),
\end{equation}
where $f_T,$ and $f_{TT}$  denote one and two times derivative with respect to the torsion scalar $T$ respectively, $\rho_m$ and $p_m$ are energy density and pressure of the matter inside the universe respectively.Furthermore, $\rho_T$ and $p_T$ stand for the torsion contribution to the energy density and pressure respectively.\\
In the Friedmann- Lemaitre-Robertson-Walker ( FLRW) background, the effective equation of state (EoS) for the universe is given by {\citep{frw}}
\begin{eqnarray}\label{14f}
\omega_{eff}=\frac{p_{eff}}{\rho_{eff}}=-1-\frac{2\dot{H}}{3H^2}\cr
\rho_{eff}=\frac{3H^2}{k^2}\cr
p_{eff}=-\frac{2\dot{H}+3H^2}{k^2},
\end{eqnarray}
 where, $\rho_{eff}$, and $ p_{eff}$ are the total energy density and pressure of the universe, respectively. When the energy density of dark energy becomes  completely dominant over the matter, one could consider $\omega_{eff}\approx \omega_{DE}=\omega_T$ and  the equation of state (EoS)parameter for such universe is given as following {\citep{kaz}}
\begin{eqnarray}\label{15}
\omega_T= -\frac{\Big[4(1-f_T-2Tf_{TT})\dot{H}+(-T-f+2Tf_T)\Big]}{2Tf_T-f-T}.
\end{eqnarray}
From Eqs (\ref{13}, \ref{14}), we have {\citep{kaz}}
\begin{eqnarray}\label{15f}
p_{DE}=-\rho_{DE}+g(H,\dot{H}),
\end{eqnarray}
where
\begin{eqnarray}\label{b17}
g(H, \dot{H})=-\frac{1}{k^2}\Big[2(1-f_T-2Tf_{TT})\dot{H} \Big].
\end{eqnarray}
It obtain from (\ref{14f}) that $p_{eff}=-\rho_{eff}-2\dot{H}/k^2$. Comparing this equation with (\ref{15f}) give{\citep{kaz}}
\begin{equation}\label{16f}
\dot{H}+\frac{k^2}{2}g(H, \dot{H})=0.
\end{equation}

The substitution of Eq. (\ref{b17}) into Eq. (\ref{16f}) gives
\begin{equation}\label{18f}
\dot{H}\Big(f_T+2Tf_{TT} \Big)=0,
\end{equation}
Since $\dot{H}\neq0$ hence  $f_T+2Tf_{TT}=0$ and the Eq.(\ref{15})reduce the following equation{\citep{kaz}}
\begin{eqnarray}\label{15f}
\omega_T=-\frac{4\dot{H}-T-f+2Tf_T}{2Tf_T-f-T}
\end{eqnarray}
At last, we  obtain the deceleration parameter to compare with the observations.
\begin{equation}\label{19}
q=-1-\frac{\dot{H}}{H^2}.
\end{equation}
Using Eq.(\ref{13}) the above equation gets
\begin{equation}\label{20}
q=1-\frac{\dot{T}}{2(-\frac{T}{6})^{3/2}}
\end{equation}
\section{Holographic $f(T)$) gravity model}

Given the fact that $f(T)$)-gravity can justify the observed acceleration of universe without any extra component as DE. This encourages us to reconstruct a $f(T)$)-gravity model according to the HDE model which has been attract a lot of interest recently. We take the HDE density as
\begin{equation}\label{21}
\rho_{DE}=3c^2H^2,
\end{equation}
where, we have set the system infrared cutoff length,$ L$, equal to the Hubble horizon $L=H^{-1}$. Using Eq.(\ref{13}) one can rewrite (\ref{21})as follows
\begin{equation}\label{22}
\rho_{DE}=-3c^2(\frac{T}{6}).
\end{equation}
From Eq.(\ref{13}, \ref{22}), i.e. $\rho_T=\rho_{DE}$, we achieve the following differential equation
\begin{equation}\label{23}
2Tf_T-f-\beta T=0,
\end{equation}
where
\begin{equation}\label{24}
\beta=1-8\pi G c^2,
\end{equation}
where it is clear that $\beta$ parameter is smaller of one i.e $\beta\rightarrow 1$.
Solving Eq. (\ref{23})given as
\begin{equation}\label{25}
$f(T)$)=T\beta+\sqrt{-T}\alpha,
\end{equation}
which is the $f(T)$)- gravity corresponding to the HDE model and $\alpha$ is an integration constant that can be obtained from a boundary condition. Recovering the present value of Newtonian gravitational constant, according to {\citep{scv}}
we need to have
\begin{equation}\label{26}
f_T(T_0)=1,
\end{equation}
where $T_0=-6H^2_0$ is the torsion scalar at the present time. Hence, using the above boundary condition and Eq. (\ref{25}) one can get
\begin{equation}\label{27}
\alpha=2\sqrt{-T_0}(\beta-1).
\end{equation}
Substituting the above equation into Eq. (\ref{25}) we achieve
\begin{equation}\label{28}
f(T)=\beta T+2\sqrt{TT_0}(\beta-1).
\end{equation}
Applying Eq. (\ref{28})into Eq. (\ref{12}) one can get the present value of $\beta$ parameter as following
\begin{equation}\label{29}
\beta=\Omega_{m_0},
\end{equation}
where $\Omega_{m_0}$ is the dimensionless quantity of the parameter density and the index $0$ indicate  the present value  quantity that is $\beta=0.26$.
It is noticeable that the holographic $f(T)$)-gravity (\ref{28}) approximately satisfies the condition $\lim_{T \to \infty}f/T\rightarrow 1$ at high redshift that is as follows
\begin{equation}\label{30}
\lim_{T\to \infty}f/T\rightarrow \beta,
\end{equation}
 where $\beta$ parameter is equal to Eq. (\ref{24})  compatible with the primordial nucleosynthesis and CMB constraints {\citep{ev,ka}}. The evolution of holographic $f(T)$)-gravity model Eq.(\ref{28}) versus $T$ is illustrated in Fig. $1$. It illustrate that the magnitude of $f(T)$ model increases with $T$ increase.
\begin{figure}[h]
\centerline{\includegraphics[width=0.43\textwidth]{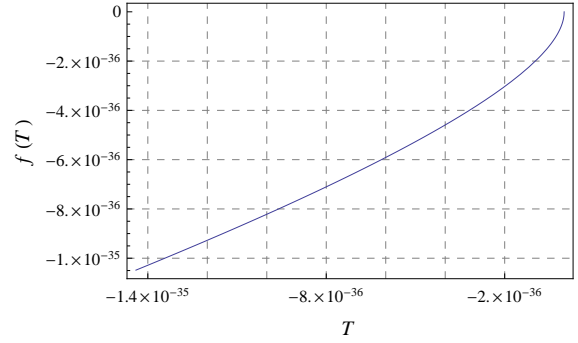}}
\caption{{\small\label{fig:F1}The plot shows the evolution of holographic $f(T)$)-gravity,Eq.(\ref{28}), versus $T$. The auxiliary parameters are $H_0=71.8KmS^{-1}Mpc^{-1}$ and $\Omega_{m_0}=0.26$ {\citep{scv}} that based on these values, the current value of $\beta=0.26,\, T_0=144\times 10^{-38}S^{-2}$
.}}
\end{figure}
\newpage

To survey the time evolution of EoS parameter, we consider the case in which the scale factor  $a(t)$ is represented by {\citep{sno,hms}}
\begin{eqnarray}\label{31}
a(t)&=&a_0(t_s-t)^{-n},\,n>0\cr
H&=&\frac{n}{ts-t},
\end{eqnarray}
Eq. (\ref{31}) explain a super accelerated FRW universe, $\dot{H}>0$, with  a Big Rip singularity at $t=t_s$.
Using Eq.(\ref{31}) one can write Eq.(\ref{13}) as
\begin{equation}\label{32}
T=-6(\frac{n}{t_s-t})^,
\end{equation}
where we get $T=-\infty$ at $t=t_s$. Figure $2$ illustrate the time evolution of the fractional torsion scalar $T t_s^2$ for four different value of $n=2,\, 3,\,4,\,5$ that have been shown with colours Blue, Red,Green and Gray respectively in the plot. It clears that $T\times t_s^2 $ decreases with increasing the time and the slope becomes sharper, also,the curve is shifted to the smaller value of $T t^2$  with increasing $n$ as can be seen from the diagram.
\begin{figure}[h]
\centerline{\includegraphics[width=0.43\textwidth]{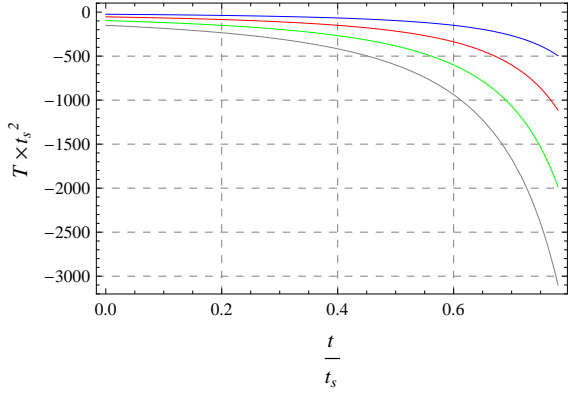}}
\caption{{\small\label{fig:F1}The plot shows the  time evolution of the  torsion scalar times $t_s^2$, Eq.(\ref{32}), versus $\frac{t}{t_s}$ for four different value of $n=2,\, 3,\,4,\,5$ that have been shown with colours Blue, Red,Green and Gray respectively.}}
\end{figure}
\newpage
Using Eq.(\ref{31}, \ref{28})and Eq.(\ref{15})the EoS parameter get as following
\begin{equation}\label{33}
\omega=-1+\frac{2}{3n(\beta-1)}.
\end{equation}
The  evolution of the EoS parameter (\ref{33})  versus $n$ parameter is illustrated in Fig. 3 .
.It shows that $\omega<-1$ is always  therefore it is in the phantom phase and in the $n=5$ we have $\omega\rightarrow -1$ which acts like $\Lambda$CDM

\begin{figure}[h]
\centerline{\includegraphics[width=0.43\textwidth]{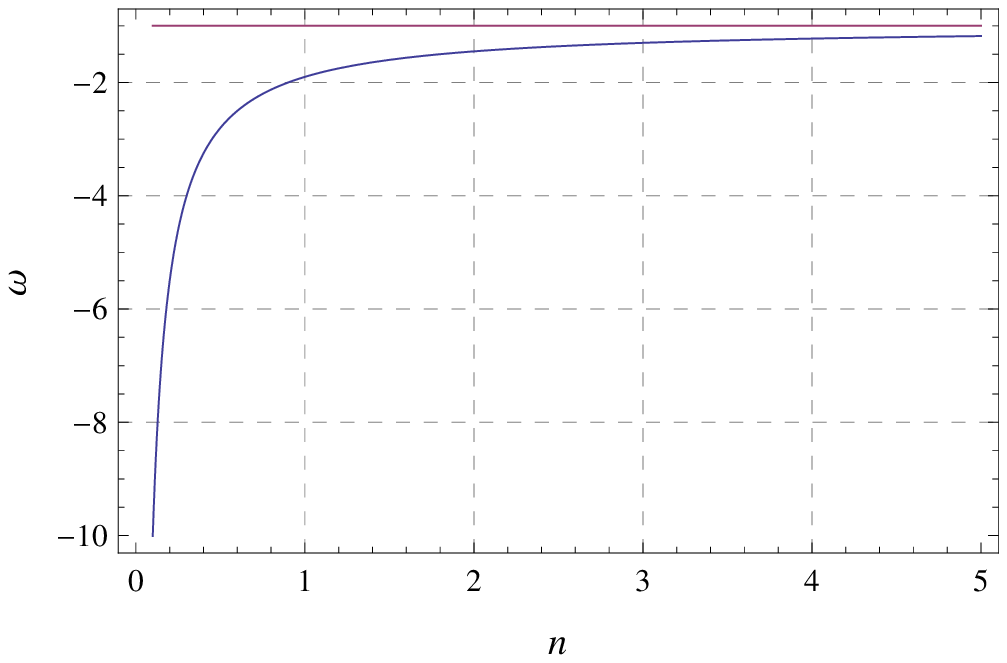}}
\caption{{\small \label{fig:F3}The plot shows the  time evolution of the EoS parameter (\ref{33}), versus $n$.}}
\end{figure}
Combining Eq.(\ref{32}) and (\ref{20}) the deceleration  parameter get
\begin{equation}\label{34}
q=-1-\frac{1}{n}.
\end{equation}
The  evolution of the deceleration  parameter (\ref{34}) is illustrated   versus $n$ parameter in Fig. 4.It clear that in the  $n\gg$ we have $q\rightarrow -1 $ which behaves like de Sitter universe. Fig. 4 shows that  the condition of the accelerated expansion is consistently  satisfied.
\begin{figure}[h]
\centerline{\includegraphics[width=0.43\textwidth]{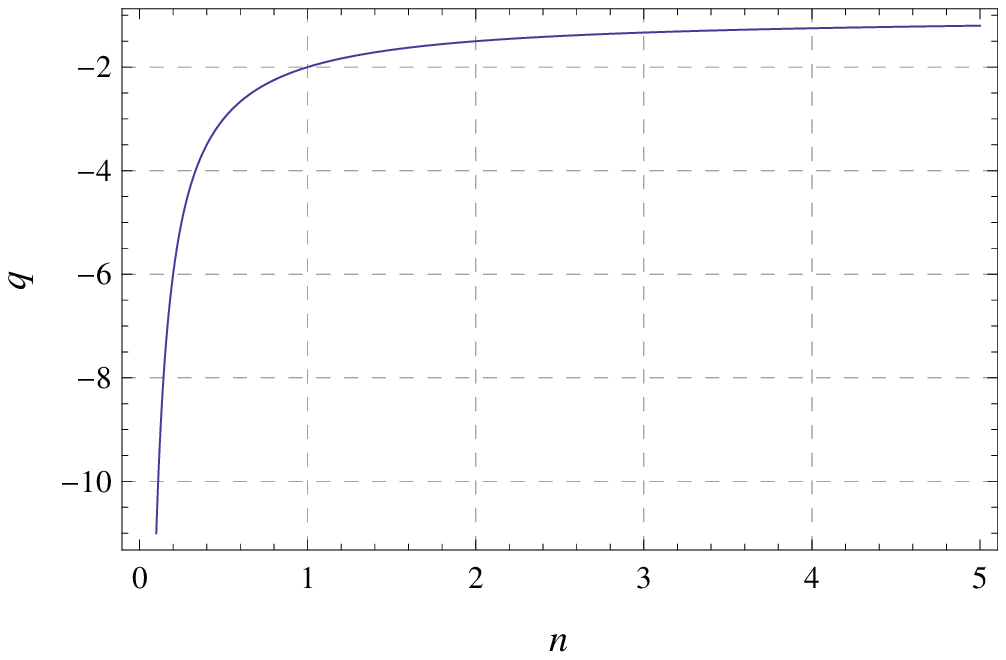}}
\caption{{\small\label{fig:F3}The plot shows the  time evolution of the EoS parameter (\ref{33}), versus $n$.}}
\end{figure}

\section{Conclusion}
We studied the corresponding relation between  $f(T)$ gravity and holographic dark energy. A kind of  energy density  from $f(T)$ have been introduced which has  the same role as the holographic dark energy density. At time future that the energy density of dark energy becomes  completely dominant over the matter, we have token   $\omega_{eff}\approx \omega_{DE}=\omega_T$. Then,   a $f(T)$ model according to the HDE scenario have been reconstructed.we concluded  that condition $f/T\rightarrow 1$ is estimated at high redshift $(T\rightarrow \infty)$ which is in agree with the primordial nucleosynthesis and CMD constraints. A power-law model for the scale factor with respect time have been taken according\cite{sno} then  the time evolution of the torsion scalar $T$  have been obtained. In addition, we calculated the effective EoS and deceleration parameters of the $f(T)$) gravity model. our results are summarized as follows:
\begin{itemize}
  \item (i) The EoS parameter $\omega_T<-1$, i.e the $\omega$ is in the phantom phase and in the $n=5$  it get $\omega\rightarrow -1$ which acts like $\Lambda$CDM.
  \item  The  $q<0$  satisfied   the condition of the accelerated expansion at the late time. Fig. 4 shows that in the  $n\gg$ it get $q\rightarrow -1 $ which behaves like de Sitter universe.
      \end{itemize}
At last, we have referred that the dynamics of the universe described by the HDE in the Einstein gravity can be explained by $f(T)$ theory without any requirement to the DE model.


\end{document}